\documentclass[runningheads]{llncs}

\usepackage[T1]{fontenc}
\usepackage[most]{tcolorbox}
\usepackage{minted}
\usepackage{hyperref}
\usepackage{graphicx}
\begin{document}
\title{Practical Source Code Recovery from Binary Functions Using Anchor-Based Retrieval and LLM Reasoning}
\titlerunning{Binary-to-Source Recovery via Anchors and LLMs}
%
\author{Charles Edward Gagnon\inst{1}\orcidID{0009-0008-4647-4398} \and
Steven H. H. Ding\inst{1}\orcidID{0000-0003-4513-200X} \and
Philippe Charland\inst{2}\orcidID{0000-0003-4051-9942} \and
Benjamin C. M. Fung\inst{1}\orcidID{0000-0001-8423-2906}
}
\authorrunning{C. Gagnon et al.}
%
\institute{McGill University, Montreal, QC H3A 0G4, Canada\
\email{charles.e.gagnon@mail.mcgill.ca, \{steven.h.ding, ben.fung\}@mcgill.ca} \and
Defence Research and Development Canada\ }
\maketitle              
\begin{abstract}
We present a practical pipeline for recovering source code from stripped binary functions by combining reverse engineering, anchor-based source code retrieval, and large language model reasoning. Our binary-to-source-code retrieval method attempts to identify the source function from a source code database, rather than generating approximate decompiled pseudocode. It extracts anchors such as strings, constants, external calls, and available function names using Ghidra, retrieves candidate files via an inverted-index search database, narrows candidates to likely function snippets, and re-ranks them with a large language model (LLM) based on disassembly, decompiled code, and source metadata. Confident matches can also serve as anchors in later passes. In an evaluation backed by our high-fidelity source code database on a stripped, optimized \texttt{tcpdump} binary, our proposed binary-to-source matching method achieves 95.2\% assembly instruction coverage. Experiments on a GitHub-based retrieval database showed lower performance with 35.5\% instruction coverage on average, mainly due to retrieval misses. These results show that source-level binary recovery excels with high-quality databases and remains a useful tool in noisy environments.

\end{abstract}

\section{Introduction}

With the ever-increasing rate of software production, reverse engineering unknown executables has become a major bottleneck for cybersecurity. Organizations must analyze a growing number of binaries originating from commercial software, malware samples, firmware images, and third-party dependencies. While software development has benefited greatly from automation and advancements in tooling, the reverse engineering process remains largely dependent on expert analysts and manual investigation. As a result, the time required to understand and assess compiled software increasingly limits the speed at which security teams can investigate incidents and detect threats.

Binary code reverse engineering is a challenging task that, even today, requires heavy human intervention. No automatic reverse engineering tool exists because compilers are fundamentally irreversible functions ~\cite{compiler-theory}. A compiler strictly maintains the semantic rules of the underlying source code, but removes all other helpful information. With optimizations enabled, a compiler will liberally get rid of object identifiers, control-flow structures, logical ordering, data structures, and even whole functions. As such, even a perfect system cannot recover as much information from the compiled binary as what was initially present in the source code. There exists a variety of tools that try to automatically obtain pseudocode from binary code. These are usually categorized as decompilers ~\cite{decompiler-theory}. However, more than a decompiler is needed to build a deep understanding of unknown software. The decompiled output represents the semantic meaning of the software, but lacks the structure, comments, and identifiers used by the developer to make sense of the code.

In this work, we present a method that fully recovers the source code of unknown binary functions. Our approach takes advantage of the significant presence of open source software in modern applications \cite{linux2022,opensource}. Our method is practical because it readily supports massive open databases such as the GitHub API and does not require the maintenance of custom machine learning models.

\section{Related Works}

\textbf{Binary to source code matching.} Numerous methods that match binary to source code already exist. A recurring theme is the use of disassembler such as Ghidra \cite{ghidra} or IDA Pro \cite{ida} to extract binary function features. One of the first contributions to this field is RESource \cite{RESource}. This method uses the extracted features to perform web queries. The query responses are parsed and injected as code comments into the disassembled binary source. BinPro \cite{BinPro} brought significant improvements by introducing machine learning in the feature extraction and matching steps. The weights associated with each feature are learned rather than hard-coded, and the queries are performed in dedicated source code databases rather than on the internet. B2SFinder \cite{B2SFinder} improved generalizability by extracting a wider range of features, rather than only focusing on string constants. These include numeric constants, exported symbols, function names, enumeration types and conditional branching. Many other methods branded under library usage detection \cite{BinProv,LibvDiff} or software bill of materials (SBOM) generation \cite{sbom-thesis,0shot-sbom,ers0} perform a very similar task. Instead of matching a specific source file to a binary routine, these methods only determine whether a known library is used by the binary code. The granularity of these methods is a common limitation. In binary to source code matching, most methods only report whether a library or source file was used, without pointing to the exact source code snippet. Furthermore, most algorithms cannot provide a rationale behind the matching outputs.

\textbf{Binary to binary matching.} A related area is binary code to binary code matching, also known as binary code similarity detection. This area of research attempts to find the closest matching binary function in a database, rather than filtering through a source code database. Current state-of-the-art methods make heavy use of machine learning and natural language processing to achieve high accuracy ~\cite{CLAP,SAFE,PalmTree,Asm2Vec,OrderMatters}. Similar to our proposed method, some directly use LLMs to perform feature extraction ~\cite{bcsd-llm}, feature analysis ~\cite{beyond-emb}, or candidate selection ~\cite{cc2}. Other methods use the source code as part of model training, and can determine the similarity between the binary source and a descriptive label of the code \cite{CLAP}. These approaches are limited by the expressiveness of the matched binary code. If the source code corresponding to the matched binary function is unknown, then identifying a similar binary function does not substantially advance the reverse engineering task. Another limitation is the capability of the model to match code compiled for different architectures and with varying optimization levels.

\textbf{Decompilation.} Decompilation tools also perform binary code to source code matching, but do so without the help of a source code database. These methods perform static analysis of disassembled binary code and generate pseudocode as output. Recent research shows that LLMs can significantly enhance traditional template-based decompilation pipelines. Models trained specifically on parallel binary-to-source corpora \cite{llm4decompile} leverage natural language capabilities to predict high-level variable names and generate more readable pseudocode than heuristic-based decompilers. A recurring limitation is that LLM-based decompilers are non-deterministic, which can make the output unreliable and hard to verify.

\begin{figure}
\centering
\includegraphics[width=\linewidth]{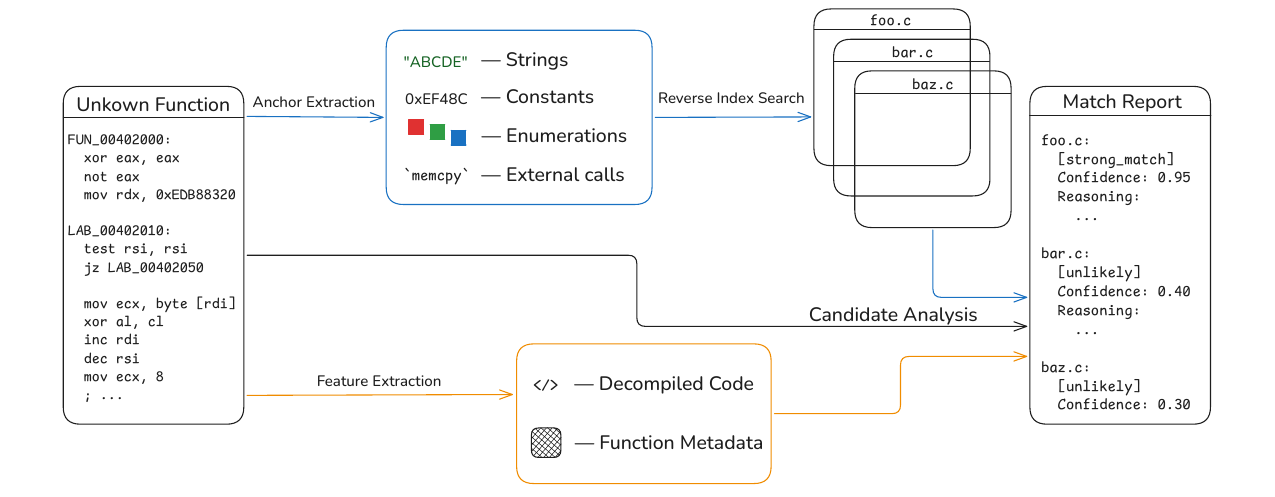}
\caption{Overview of our method. An unknown function is first disassembled and its features are extracted. The anchors are used to perform the initial database query. The disassembled code, along with information-dense features, is used to re-order the top-\(k\) candidates. In our experiments, \(k\) is set to 12.}
\label{fig:overview}
\end{figure}

\section{Methodology}

We formulate binary-to-source recovery as a retrieval and verification problem. Given a stripped binary function \(b\), the goal is to identify the source-level function \(s\) from a source code database \(D\) that most likely produced \(b\) after compilation. Unlike decompilation, which attempts to synthesize readable pseudocode from binary instructions alone, our method searches for an existing source implementation and verifies whether it corresponds to the observed binary routine. Our method assumes that the target source code, or a sufficiently similar implementation, is present in the backing database. It does not expect the presence of debug symbols, although it can use them if available.

Our pipeline is shown in \autoref{fig:overview}. First, static analysis is applied to the target binary function to extract compact retrieval features and denser function features. Second, retrieval features are used as anchors to retrieve candidate source files from a source code database. Third, the retrieved files are narrowed to function snippets and ranked based on the number of anchors matched. Finally, a large language model verifies the most promising candidates and re-ranks them by comparing the dense function features with each candidate source function. Confident matches can then be propagated through the binary call graph as additional anchors for later passes.

\subsection{Feature Extraction}

The first stage extracts features from the target binary function using Ghidra~\cite{ghidra}. We distinguish between two classes of features: \textit{anchors}, which are used for retrieval, and dense features, which are reserved for later verification and re-ranking.

Anchors are features used in the initial source code database query to find relevant source files. They include strings and numerical constants, external calls, and function names. The most reliable anchor is a string constant. It provides heavy filtering capabilities and a very low false positive rate. Numerical constants are also useful, but their smaller domain makes them more susceptible to false positives. They are especially useful in functions that use unique values, such as hashing routines and other cryptographic primitives. External function calls are less unique but remain useful when combined with others. These are usually \texttt{libc} function calls or compiler builtins, such as \texttt{memset} or \texttt{memcpy}. Lastly, internal function names are highly effective when available, although they are usually absent from stripped production binaries.

Dense features are not used directly in the initial database query. Instead, they are used during candidate verification. These include the disassembled instructions, the decompiled pseudocode, and metadata recovered by Ghidra. These features provide a richer description of the binary function, but they are too verbose to be used directly as inverted-index search keys.

\subsection{Source Code Retrieval}

The second stage uses the extracted anchors to retrieve candidate source files from the backing database. To remain scalable, the database is accessed through an inverted-index search interface. The query consists of the anchors extracted from the binary function and the database returns source files ranked according to their textual relevance to those anchors. This abstraction allows the method to operate over different source code databases. In a controlled enterprise setting, the database can be implemented as a domain-specific and curated index of source files. In a large-scale real-world setting, the same interface can be implemented using external code search services such as the GitHub API. The retrieval stage does not depend on a specific database implementation, but its effectiveness depends on the completeness, ranking quality, and noise level of the backing corpus.

\subsection{Candidate localization}

Once matching source files are found, the code corpus is reduced to only contain relevant snippets. Our method uses a large combination of techniques to handle a number of issues. The first issue comes with constants and preprocessor macros. Standard code style guidelines~\cite{barr2018} recommend constants to be declared before their use, usually at the very beginning of a file. This poses an issue because naive snippet extraction would very often return inconsequential portions of the source file. To alleviate this issue, a preprocessor analyzes the statement containing the anchor and decides whether this statement constitutes a definition or a use site. If the anchor is found at a definition site, all use sites of the identifier to which the constant is assigned are added to the candidates, and the definition site is ignored. Another issue stemming from compiler optimizations is function inlining. Anchors can point to a function that is not actually part of the compiled binary because it was inlined in the body of the calling function. As such, our method uses heuristics to determine whether a function is likely to be inlined by the compiler (function size, presence of inline tags, and control flow complexity). If the target function is deemed likely to be inlined, the function name is added to the set of anchors. This largely increases the effectiveness of our method. It helps verification to find the full source code function linked to the compiled binary routine, rather than a smaller function that was inlined. With these techniques, our method is able to reliably find source snippets that relate to the anchors. Snippets use function boundary detection to fit whole functions as the result. The number of anchors in the snippet and the ratio of anchors to lines of source are used to rank source snippets in order of importance.

\subsection{Function similarity reasoning}

\begin{figure}[htbp]
\centering

\begin{minipage}[t]{0.48\textwidth}
\textbf{Decompiled code}
\begin{minted}[fontsize=\scriptsize,breaklines]{c}
undefined8 FUN_0010bd40(undefined8 *param_1,long param_2)
{
char *__file;
undefined8 uVar1;
int iVar2;
uint uVar3;
FILE *pFVar4;
int *piVar5;
char *pcVar6;
size_t __n;
uint *puVar7;
int iVar8;

if ((*(int )(param_2 + 0x500) == 2) ||
(((int )(param_2 + 0x500) == 0 && ((char *)((long)param_1 + 9) == '\0')))) {
pFVar4 = fopen(__file,"wb");
if (pFVar4 != (FILE *)0x0) {
LAB_0010be15:
param_1[2] = pFVar4;
(undefined1 () [16])(param_1 + 3) = (undefined1  [16])0x0;
*(undefined2 )((long)param_1 + 10) = 0x101;
return 1;
}
}
else {
/* … */
}

/* … */
piVar5 = __errno_location();
pcVar6 = strerror(*piVar5);
FUN_00115b40(uVar1,"Failed to open the file %s: %s",__file,pcVar6);
return 0;
}
\end{minted}
\end{minipage}
\hfill
\begin{minipage}[t]{0.48\textwidth}
\textbf{Source code}
\begin{minted}[fontsize=\scriptsize,breaklines]{c}
bool tool_create_output_file(
struct OutStruct *outs,
struct OperationConfig *config)
{
FILE *file = NULL;
const char *fname = outs->filename;
DEBUGASSERT(outs);
DEBUGASSERT(config);
DEBUGASSERT(fname && *fname);

if(config->file_clobber_mode == CLOBBER_ALWAYS ||
(config->file_clobber_mode == CLOBBER_DEFAULT &&
!outs->is_cd_filename)) {
/* open file for writing */
file = curlx_fopen(fname, "wb");
}
else {
int fd;
do {
/* … */
}
}

if(!file) {
char errbuf[STRERROR_LEN];
warnf("Failed to open the file %s: %s", fname,
curlx_strerror(errno, errbuf, sizeof(errbuf)));
return FALSE;
}
outs->regular_file = TRUE;
outs->fopened = TRUE;
outs->stream = file;
outs->bytes = 0;
outs->init = 0;
return TRUE;
}
\end{minted}
\end{minipage}

\vspace{1.5em}

\begin{tcolorbox}[colback=gray!10,colframe=gray!20,coltitle=black]
The candidate source in \texttt{curl/src/tool\_cb\_wrt.c} matches the decompiled function closely in control flow, error handling, and key string literals such as \texttt{“Remote filename has no length”}, \texttt{“out of memory”}, and \texttt{“Failed to open the file \%s: \%s”}. The source function \texttt{tool\_create\_output\_file} implements the same logic of opening files with conditions on file existence, modes, and error reporting. The use of \texttt{open}, \texttt{fopen}, \texttt{fdopen}, \texttt{close}, \texttt{malloc}, \texttt{memcpy}, \texttt{curl\_msnprintf}, and \texttt{errno} handling matches the decompiled code’s calls and flow. The numeric constants and offsets (e.g., \texttt{0x508}, \texttt{0x500}) correspond to field accesses in the source. The source function signature and parameter usage align well with the decompiled code’s \texttt{param\_1} and \texttt{param\_2} usage. This is a direct match to the decompiled function, not just a helper or macro.
\end{tcolorbox}

\caption{Example of successful retrieval from decompiled code to source code. The language model is able to precisely explain the reasoning behind the match.}
\label{fig:code_comparison}
\end{figure}

The novelty of our method comes from the way in which it handles source snippet candidates. As highlighted in \autoref{fig:code_comparison}, instead of relying on a simple ranking based on anchor density or other heuristics, our method uses the capabilities of LLMs to perform an in-depth analysis of each candidate. The representations extracted from the binary function are provided in the language model query. The response follows a specific schema, reporting the matching \textbf{score}, the \textbf{reasoning} explaining which elements of the function match and differ, and a final \textbf{verdict} that indicates the confidence of the model in the match.

A surprising benefit is that our method can also be applied iteratively. Once a binary function is confidently matched to a source function, the recovered source identity can serve as a new anchor for caller functions. For example, if a binary function at address \(x\) is matched to a source function \(f\), then calls to \(x\) from other binary functions can be interpreted as calls to \(f\). The name \(f\) can then be added to the anchor set of those caller functions. This secondary-pass mechanism allows information recovered from high-confidence matches to propagate through the call graph. It is especially useful for functions that contain few intrinsic anchors but call functions that have already been identified. This way, the method can recover additional source functions that would be difficult to retrieve from local features alone.

\section{Experiments}

To assess the capabilities of our method, we first experimented it in a controlled environment against our own database of open source projects. We then performed an experiment on the public GitHub API, showcasing how our method generalizes to any database already in production. The metrics presented for each experiment include the percentage of functions correctly identified in first place (Hit @ 1) and in the top three (Hit @ 3). We additionally provide the mean reciprocal rank (MRR)
\[
\mathrm{MRR} = \frac{1}{|B|} \sum_{i=1}^{|B|} \frac{1}{\mathrm{rank}_i}
\]
where \(B\) is the set of binary functions. Finally, we report the percentage of the binary that is fully reverse engineered by our tool (Coverage). That is, the percentage of assembly instructions that are part of a binary function that is correctly mapped to its source code function.

\subsection{Controlled Environment Evaluation}

For our first experiment, the \texttt{tcpdump} binary is reverse engineered with our approach against a private database containing source files from \(8,061\) open source projects. The \texttt{tcpdump} binary is compiled with optimization level O2, for the \texttt{x86\_64} architecture. A map between function addresses and their source is available as ground truth. To do so, the binary is first compiled with debug symbols. An address-to-function map that stores the original function name, the source file and line number is extracted from the binary using the \texttt{nm} tool~\cite{gnu_nm}. Finally, the binary is stripped of all debug symbols and passed to our tool for analysis. Results are reconstructed by comparing the output of our tool with the address-to-function map. If function names match and both functions are located in the same source file, then the function is considered correctly mapped. We selected GPT-5.5 for candidate verification in this experiment, seeking to obtain results that are not limited by language model capabilities.

\begin{table}[htbp]
\centering
\caption{Evaluation metrics of our experiment on the \texttt{tcpdump} binary with a curated database.}
\label{tab:evaluation_metrics_basic}

\begin{tabular}{|l|@{\hspace{25pt}}c@{\hspace{25pt}}|}
\hline
\textbf{Metric} & \textbf{Value} \\ \hline
Queries Evaluated & 815 \\ \hline
Queries with Hit  & 714 \\ \hline
Miss              & 101 \\ \hline
Top 1             & 698 \\ \hline
Top 3             & 714 \\ \hline
Hit@1             & 0.8564 \\ \hline
Hit@3             & 0.8761 \\ \hline
MRR               & 0.8660 \\ \hline
Coverage          & 95.2\% \\ \hline
\end{tabular}
\end{table}

As evident from \autoref{tab:evaluation_metrics_basic}, our method performs remarkably well on classical reverse engineering of code compiled from \texttt{C++}. Our method alone covers 95\% of all instructions in the binary with a correctly matched source function. Most failed matches stem from very small assembly routines that do not contain any reliable anchor for the initial database query to be effective. For instance, 101 functions from the \texttt{tcpdump} binary contained zero anchors, all of which could not return any meaningful results from the database. However, out of the 714 queries that did contain anchors, all valid matches were found in the top 3 candidates from the language model analysis. This is explained by the database containing very little noise (such as markdown documents, configuration files and logs). As such, queries are efficient, especially when unique anchors are provided. It is worth noting that with an unknown malware sample or proprietary binaries, our method is unable to reverse engineer functionality that is not previously found in the database. Yet, it is known that a vast majority of proprietary software constitutes code reuse of open source libraries ~\cite{linux2022,opensource}. Nevertheless, our method can still find close matches when the exact match is not found in the database, given a large enough dataset.

\subsection{Real-World Use Case}

Next, we evaluate the method in a real-world setting by replacing the curated database with the public GitHub code search interface. This experiment is intended to measure how the pipeline behaves when the backing database is large, noisy, and externally ranked. The set of binaries evaluated in this experiment consists of ubiquitous open source utilities: \texttt{BusyBox}, \texttt{curl},  \texttt{sqlite3}, and \texttt{dropbear}. All binaries are compiled for Linux \texttt{x86\_64} with optimization level O2. We also vastly reduce the size of the language model, going from GPT 5.5 (estimated 9.7T parameters~\cite{modelestimate}) to GPT 5 mini (estimated 410B parameters~\cite{modelestimate}).

\begin{table}
\caption{Evaluation metrics of our experiment using the GitHub API as the backing source code database.}\label{tab:github}
\centering
\begin{tabular}{l|@{\hspace{10pt}}cccc@{\hspace{10pt}}|c}
\hline  \hline
& \texttt{BusyBox} & \texttt{sqlite3} & \texttt{curl}   & \texttt{dropbear} & average \\ \hline
Hits @ 1               & 0.134   & 0.191   & 0.541  & 0.265    & 0.283   \\
Hits @ 3               & 0.191   & 0.268   & 0.712  & 0.354    & 0.381   \\
Instruction Coverage   & 18.7\%  & 26.0\%  & 67.1\% & 30.4\%   & 35.5\%  \\
False Positives        & 0.068   & 0.140   & 0.329  & 0.159    & 0.174   \\ \hline
Database Miss          & 0.751   & 0.668   & 0.130  & 0.577    & 0.531   \\
Hits @ 1 among retrieved & 0.536   & 0.577   & 0.622  & 0.625    & 0.590   \\
Hits @ 1 no re-raking & 0.080 & 0.068 & 0.336 & 0.245 & 0.182 \\ \hline
Total                  & 4182    & 1867    & 146    & 359      &         \\ \hline \hline
\end{tabular}
\end{table}

The results in \autoref{tab:github} are substantially weaker than those obtained in the controlled experiment. Across all four binaries, the average Hits @ 1 drops to 0.283 and the average instruction coverage drops to 35.5\%. This reduction is expected. The GitHub API does not provide a clean, deduplicated corpus of source files, unlike our curated database. Instead, the retrieval stage must operate over a large and heterogeneous source code collection whose ranking is out of our control. As a result, many functions fail before the language model re-ranking stage is reached.

The dominant failure mode is database miss. On average, 53.1\% of functions are classified as database misses, meaning that the correct source function was not present among the retrieved candidates. This is especially visible for \texttt{BusyBox}, \texttt{sqlite3}, and \texttt{dropbear}, where the database miss rates are 0.751, 0.668, and 0.577, respectively. These values indicate that the main bottleneck in the real-world setting is not only source-to-binary reasoning, but retrieval. If the correct file is absent from the candidate set, the re-ranker cannot recover the match. This interpretation is supported by the Hits @ 1 among retrieved metrics. When database misses are excluded, the average Hits @ 1 rises from 0.283 to 0.590. This shows that the real-world performance of the full system is tightly coupled to the stability and quality of the backing search engine. The comparison with the non-LLM baseline further clarifies the role of the language model. Without the language model re-ranker, the average Hits @ 1 is only 0.182. Adding the re-ranker increases this value to 0.283, and the improvement is especially large for \texttt{curl}, where Hits @ 1 rises from 0.336 to 0.541. Nevertheless, LLM reasoning cannot compensate for missing candidates.


\section{Limitations and Conclusion}

While our method performs excellently with a controlled data source and a performant model, the output is not as ideal in a noisy environment. First, having a very large data source such as the one provided by GitHub increases the potential of not finding the correct match in the first dozen results from the database. Furthermore, the GitHub database contains a lot of duplicates for popular projects because of repository forks, exacerbating the problem.

\begin{figure}
\centering

\begin{minipage}[t]{0.48\textwidth}
\begin{minted}[fontsize=\scriptsize]{c}
char *get_name(struct Person *p) {
return p->name;
}
\end{minted}
\end{minipage}
\hfill
\begin{minipage}[t]{0.48\textwidth}
\begin{minted}[fontsize=\scriptsize]{c}
undefined8 get_name(long param_1)
{
return *(undefined8 *)(param_1 + 0x08);
}
\end{minted}
\end{minipage}

\vspace{0.5em}

\begin{minipage}[t]{0.48\textwidth}
\begin{minted}[fontsize=\scriptsize]{c}
char *get_email(struct Person *p) {
return p->email;
}
\end{minted}
\end{minipage}
\hfill
\begin{minipage}[t]{0.48\textwidth}
\begin{minted}[fontsize=\scriptsize]{c}
undefined8 get_email(long param_1)
{
return *(undefined8 *)(param_1 + 0x16);
}
\end{minted}
\end{minipage}

\caption{
Comparison of two structure-field accessors and their corresponding
Ghidra decompilations. The decompiled functions are nearly identical,
differing only in the accessed field offset (\texttt{0x08} versus
\texttt{0x16}).
}
\label{fig:field-accessors}
\end{figure}

Second, our method is limited by the ambiguity of decompiled code. The language model re-ranker makes extensive use of the decompiled code to compare query results and correctly find the most relevant match. An issue arises when the decompiled code is very similar to the source function, but is not semantically identical because of subtle differences. In these circumstances, the re-ranker has the tendency to confirm the first incorrect candidate before considering the second. For instance, ~\autoref{fig:field-accessors} highlights how two functions accessing different fields of a structure can look very similar, even though the data subject is completely different. Third, our method is limited when considering functions that are represented by a limited set of anchors. Cryptographic or signal-processing kernels usually fall into this category of purely computational functions that do not interact with the operating system or use strings. The database retrieval stage struggles in such scenarios.

This work presented a practical approach for recovering source code from unknown binary functions by combining traditional reverse engineering techniques with source code search and language model-based candidate analysis. Rather than producing only decompiled pseudocode, the proposed pipeline attempts to identify the source function from a large database of available code. By extracting anchors such as strings, constants, external calls, and function names, narrowing candidate snippets, and using an LLM to reason about function similarity, the method provides a more structured and interpretable path toward binary-to-source recovery. A crucial aspect of the method is the explicit reasoning about the evidence provided by the language model, making the recovery process more interpretable by experts. The experimental results show that this approach can be highly effective when the correct source code is available in a clean and well-indexed database. In the controlled setting, the method achieved strong matching accuracy and high instruction coverage, demonstrating that the combination of anchor-based retrieval and language-model re-ranking can recover a substantial portion of a stripped binary. The GitHub-based experiment further shows that the method can operate in a realistic and noisy environment, though its performance is currently limited by duplicated repositories and retrieval quality. Future work should focus on improving the retrieval stage, expanding and deduplicating source code databases, and incorporating stronger models or additional program analysis signals to reduce false positives. Secondary passes also offer a promising direction, since confidently matched functions can become new anchors for discovering surrounding code. This work contributes a scalable framework for source-level binary recovery, showing that reverse engineering can be significantly accelerated when static analysis, large source code corpora, and language model reasoning are combined effectively.

\begin{credits} 
\subsubsection{\discintname} The authors declare that they have no competing interests.
\end{credits}
%
%
%
\bibliographystyle{splncs04}
\bibliography{references}

\end{document}